\documentclass[a4paper,11pt,one-side]{article}
\usepackage{xcolor}
\usepackage{amsmath}
\usepackage{graphicx}
\usepackage[resetlabels]{multibib}
\newcites{latex}{References for Supporting Information}

\usepackage[colorlinks=true,urlcolor=black]{hyperref}

\usepackage{authblk}

\usepackage[a4paper, total={6.5in, 9in}]{geometry}

\usepackage{chngcntr}

\author[1]{Hongcheng Tao}
\author[1]{Francesco Danzi}
\author[2]{Christian E. Silva}
\author[1]{James M. Gibert \thanks{Correspondence to: jgibert@purdue.edu}}
\affil[1]{Purdue University, 610 Purdue Mall, West Lafayette, Indiana, 47907, USA}
\affil[2]{Escuela Superior Polit\'ecnica del Litoral, ESPOL, Km 30.5 Vía Perimetral, Guayaquil, Ecuador}
\date{}
\title{Heterogeneous Digital Stiffness Programming}

\begin{document}

\maketitle


\begin{abstract}
Digital stiffness programmability is fulfilled with a heterogeneous mechanical metamaterial. The prototype consists of an elastomer matrix containing tessellations of diamond shaped cavities selectively confined with semi-rigid plastic beam inserts along their diagonals. Unit-cell perturbations by placing or removing each insert reshape the global constitutive relation whose lower and upper bounds corresponding to the configurations with all holes empty and all inserts in place, respectively, are significantly distant from each other thanks to a gap between the moduli of the elastomer and the inserts. Bidirectional operation is achieved by mixing insert orientations where longitudinal inserts enhance the macroscopic stiffness in compression and transverse ones tension. Arranged digital representations of such local insert states form the explicit encoding of global patterns so that systematic stiffness programming with minimal changes in mass is enabled both statically and in situ. These characteristics establish a new paradigm in actively tuning vibration isolation systems according to shifts in the resonance of base structures.
\end{abstract}
\section{Introduction}
Mechanical metamaterials achieve exotic physical properties beyond affine assumptions by chemical constituents through creative designs of geometry~\cite{bertoldi2017flexible,surjadi2019mechanical}. Programmable mechanical metamaterials, in particular, represent platforms for tunable global behaviors via manipulations of unit cell deformation states either a priori in the fabrication phase or through in situ external perturbations~\cite{fang2018programmable,coulais2016combinatorial,wang2014harnessing,silverberg2014using,wang2019architected,haghpanah2016multistable}. Structural multistabilities such as buckling patterns of beam arrays~\cite{florijn2014programmable,medina2020navigating,rafsanjani2015snapping,el2019exploiting}, bistable buckling domes~\cite{udani2021programmable} or Kirigami slits~\cite{yang2018multistable,an2020programmable,rafsanjani2016bistable,tang2017programmable,hwang2018tunable} are widely employed for such controllable local state switching. The joint effort of unit-cell behaviors and architecture of the assembly modifies corresponding global constitutive relations toward characteristics such as quasi-zero or negative stiffness zones, negative Poisson's ratio and nonuniform curvatures for applications in energy absorption~\cite{shan2015multistable}, vibration isolation~\cite{zhang2021tailored}, targeted shape morphing~\cite{guseinov2020programming,jin2020kirigami}, etc. The present work introduces a heterogeneous programmable mechanical metamaterial to address several common challenges in stiffness programming strategies. First, while a comprehensive range of attainable stiffness values is generally desired for efficient programming, such bounds are inevitably limited by the fundamental mechanism of a design. For example, programmable stiffnesses of a Kirigami sheet~\cite{yang2018multistable} with a constant cut pattern are bounded by two extreme configurations, where all slits buckle identically.

Heterogeneity is recognized as a natural approach to extend such range~\cite{an2020programmable} and a similar concept is exploited in the present work to implement a programmable stiffness range spanning between a lower bound corresponding to thin-walled collapsible structures and an upper bound of the constituent bulk material. Second, a programmable material typically operates in unidirectional loading and programming by perturbations on unit cells requires adequate deformation to be present, e.g., porous structures with arrays of thin beam elements~\cite{medina2020navigating} usually work in compression and the buckling states of unit cells may be perturbed only when they are already deformed, while a Kirigami sheet functions in tension and buckling of its slits may only be controlled when a load is applied. The present work adopts local geometric confinements~\cite{florijn2014programmable} as the means of perturbation for both in situ and static programming capabilities, as well as bidirectional operations utilizing different confinement configurations for compression and tension. Third, geometrically admissible patterns predicted by permutations of unit cell states can be mechanically unstable in practice, which may call for analytical or experimental efforts to distinguish. This generally results from strong structural couplings among neighboring unit cells, which also causes difficulties in the independent perturbation of each unit cell so that clusters of cells sometimes need to be manipulated simultaneously. The design to be presented aims at gaining the confidence that all combinations of unit cell states are physically reachable. This allows a natural digital encoding of all programmable patterns by arranged representations of unit cell states~\cite{haghpanah2016programmable}, which should expedite explicit modeling, analysis and optimization of programmable behaviors especially with an increased number of unit cells. Lastly, the redundancy of such encoded global patterns due to geometric and functional symmetries limits the resolution of programming, e.g., $n$ unit cells with identical contributions support $n$ possible stiffnesses~\cite{chen2021reprogrammable}.  Similarly, it will be shown that the 2-dimensional and finite nature of the present design promotes a programming resolution exponentially proportional to the number of unit cells.
 
\section{Results and Discussion}
\begin{figure}
\includegraphics[width=\linewidth]{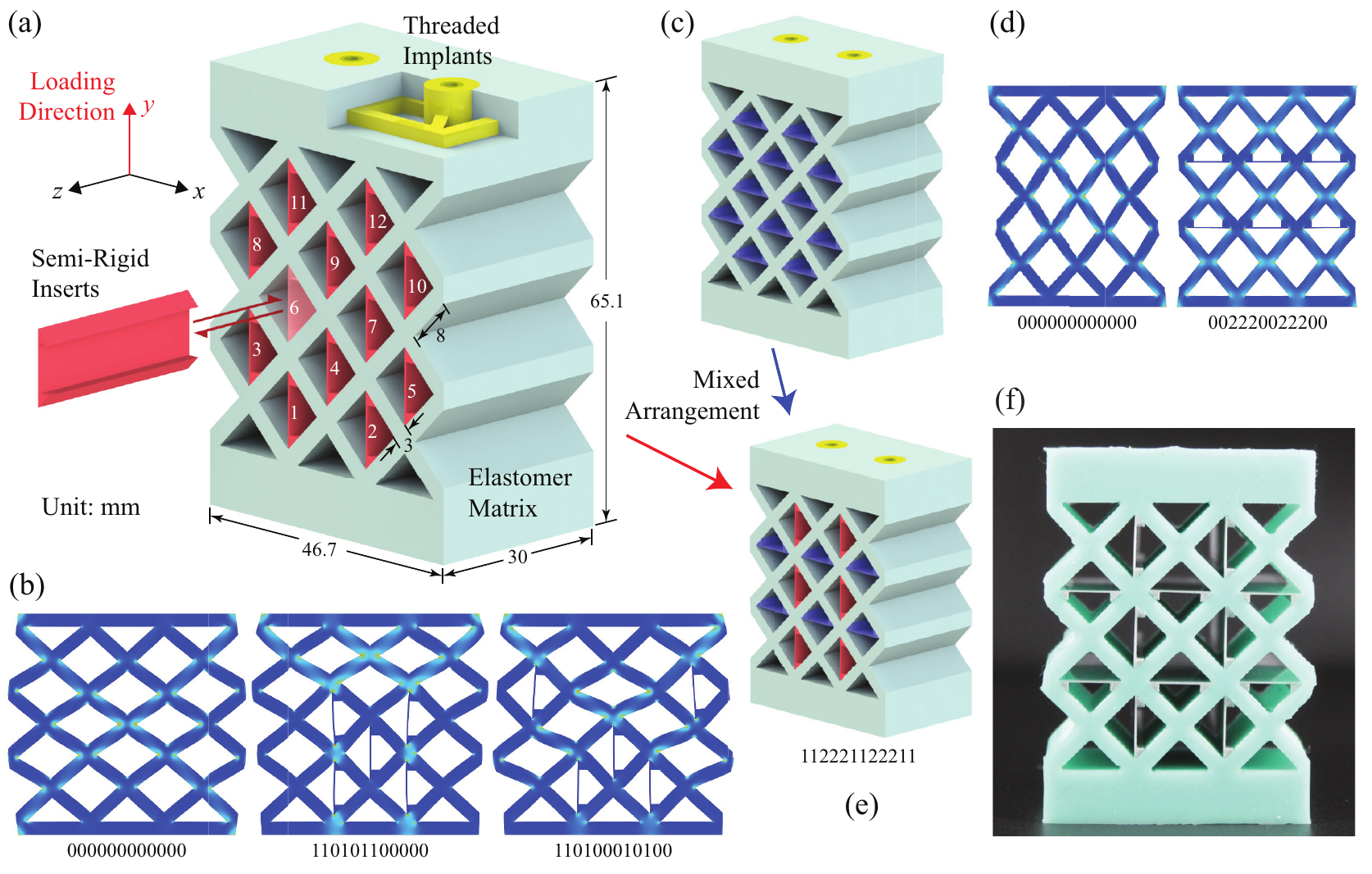}
\caption{Schematics of the heterogeneous programmable mechanical metamaterial. a) Compositions and dimensions of the prototype filled with longitudinal inserts enhancing compressive stiffness, where integers indicate the corresponding digit of each unit cell for pattern encoding. b) Linear finite element analysis for deformations of selected patterns in compression with color map indicating magnitudes of volumetric strain. c) Prototype filled with transverse inserts enhancing tensile stiffness. d) Linear finite element analysis for deformations of selected patterns in tension. e) Prototype with mixed insert orientations. f) Fabricated sample.}
\label{fig1}
\end{figure}

The heterogeneous mechanical metamaterial accommodates stiffness programming within a wide range whose lower bound is set by a 2-dimensional collapsible elastomer matrix as a thin-walled grid formed by 12 tessellated diamond shaped holes, as demonstrated in Figure~\ref{fig1}a. The empty grid deforms homogeneously under uniaxial loading, as shown in Figure~\ref{fig1}b and Figure~\ref{fig1}d, where the global strain is shared uniformly among bended elastomer walls. The matrix is cast in silicone rubber (Smooth-On Mold Star 15) with embedded threaded implants for mounting onto test fixtures. Plastic beam inserts with a width of the diamond diagonal are printed in commercial PLA (modulus over 2000 MPa) which is considered semi-rigid compared to the elastomer (modulus below 0.4 MPa). Each insert can be geometrically locked in position when slid into a diamond hole along one of the two diagonals so that when placed longitudinally (Figure~\ref{fig1}a) and transversely (Figure~\ref{fig1}c) it confines the deformation of the four surrounding elastomer walls in global compression and tension, respectively. The effective stiffness of unit cells with inserts is transformed from that of a thin-walled elastomer grid to that of the bulk elastomer around locations of contact with the inserts, where strain is highly concentrated. The addition of inserts therefore enhances the macroscopic stiffnesses correspondingly, so that the configuration where all unit cells are filled with longitudinal inserts represents the upper bound for the global constitutive relation in compression and transverse ones that for tension. Stiffness programming between such bounds can be conducted by selective insert arrangements, whose resolution (number of achievable stiffness values) is decided by the number of independent insert patterns under geometric symmetries. A mixed pattern with inserts in both directions, as depicted in Figure~\ref{fig1}e, enables bidirectional stiffness programming within a slightly narrower range. The outstanding elasticity of the silicone matrix facilitates the recovery of concentrated contact strains as well as in situ and independent placement and removal of each insert under moderate global deformations. The latter rationalizes a straightforward digital encoding scheme where each programmable insert pattern is represented by a string of integers listing the states of each unit cell with ``0" for no insert, ``1" for longitudinal insert and ``2" for transverse, which is adopted in Figure~\ref{fig1} and what follows.

\begin{figure}
  \includegraphics[width=\linewidth]{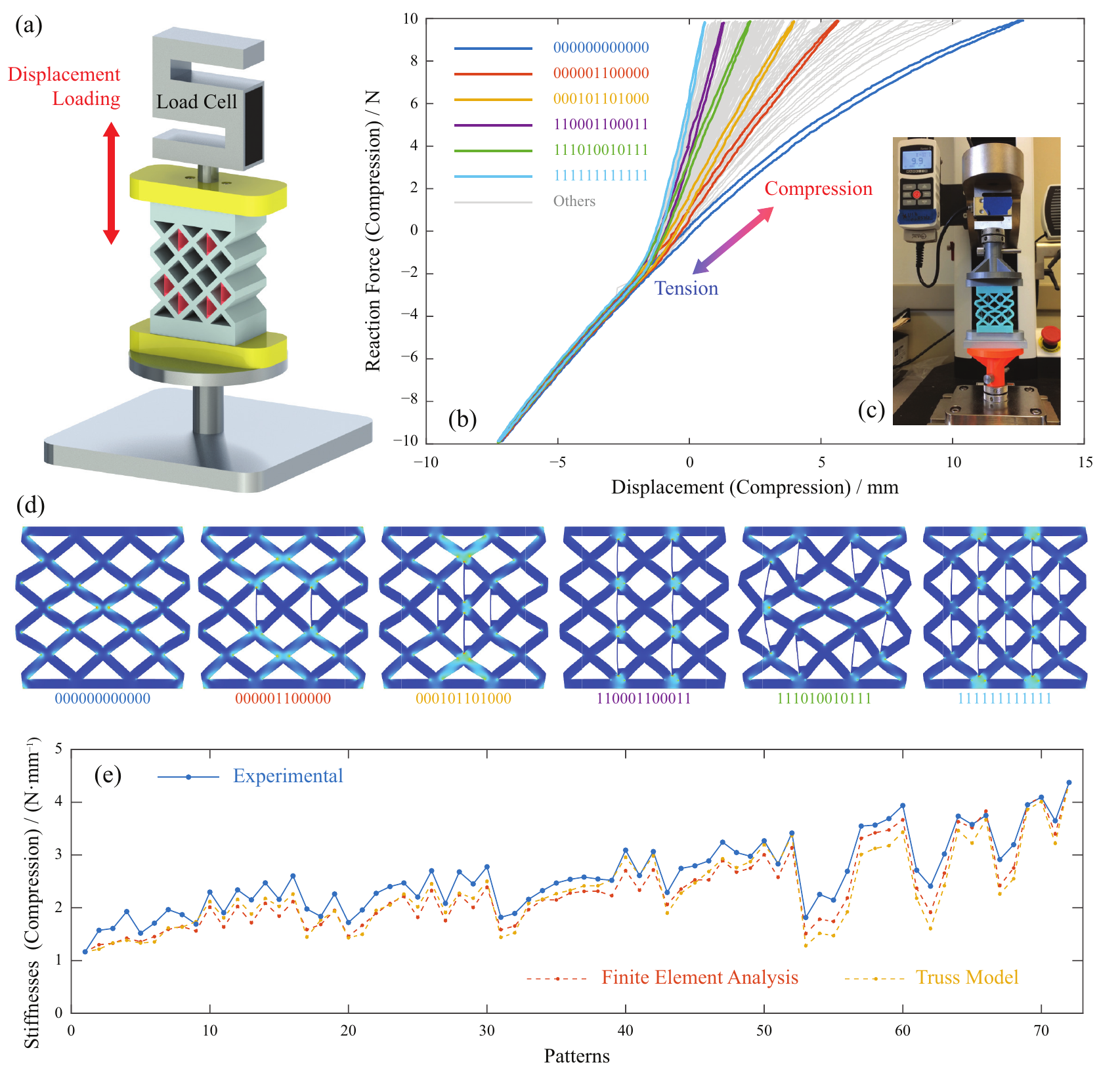}
  \caption{Compressive stiffness programming with longitudinal inserts. a) Test setup. b) Global constitutive relations for 72 independent insert patterns. c) Apparatus details. d) Linear finite element analysis for deformations of highlighted patterns in compression. e) Compressive stiffnesses of tested patterns with comparison to numerical predictions.}
  \label{fig2}
\end{figure}

\begin{figure}
  \includegraphics[width=\linewidth]{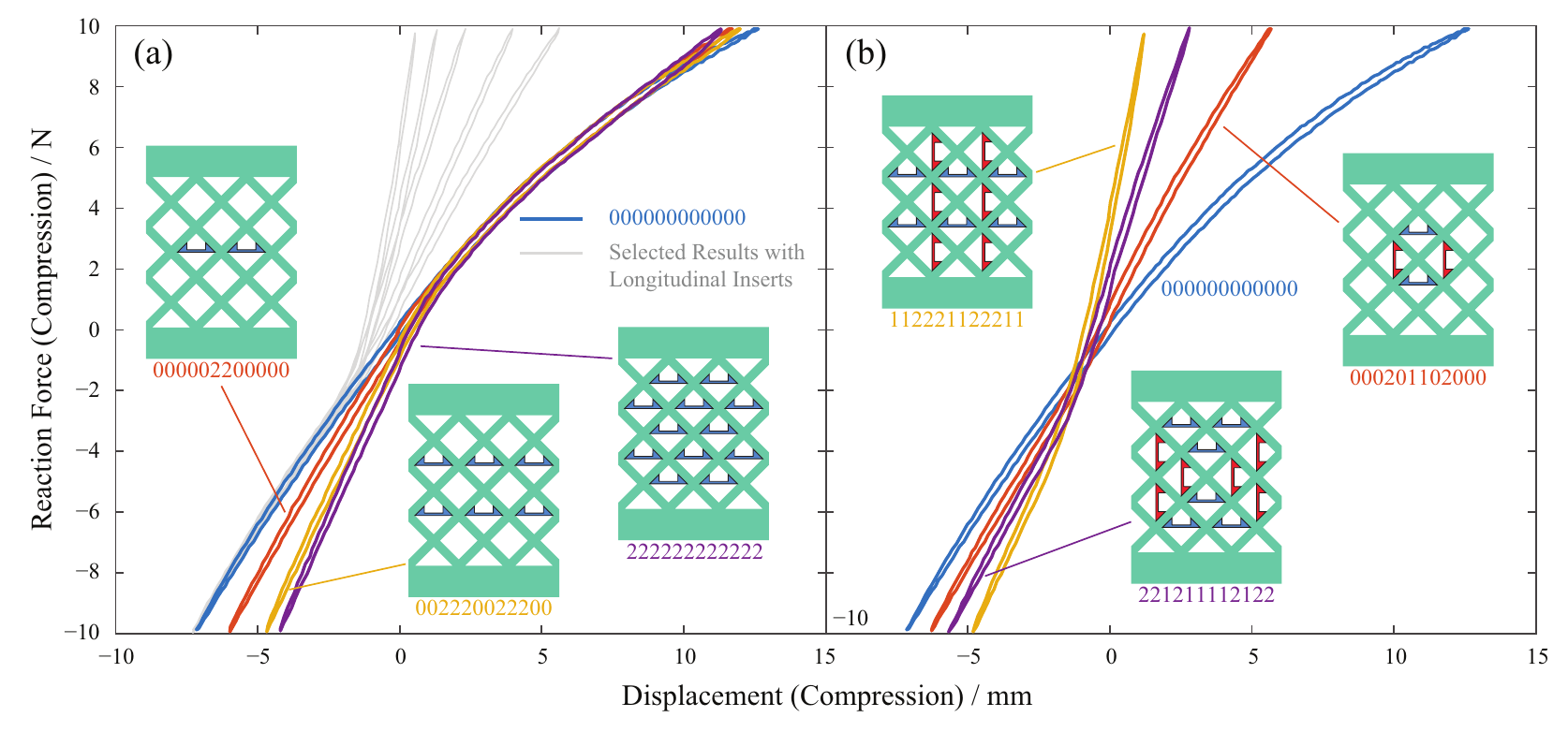}
  \caption{Tensile and bidirectional stiffness programming. Experimental global constitutive relations for a) selected transverse insert patterns, and  b) selected mixed insert patterns.}
  \label{fig3}
\end{figure}

Quasi-static stiffness programming in the linear region is examined with a universal test procedure where for each insert pattern the prototype undergoes three bidirectional loading cycles between terminations at 10 N reaction forces on a motorized compression test stand (MARK-10 ESM1500) at a rate of 15 mm/min, as outlined in Figure~\ref{fig2}a. Considering longitudinal inserts only and assuming horizontal mirror symmetry, 72 independent patterns are tested where for each the global displacement-reaction relation is extracted from the third cycle, as plotted in Figure~\ref{fig2}b. These patterns exhibit nearly identical mechanical behaviors in tension while spreading densely in the compression domain, the calculated stiffnesses of which match closely the predictions by finite element analysis in COMSOL and an theoretical truss model, both linearly rescaled with two parameters representing estimated material properties, as enumerated in Figure~\ref{fig2}e with patterns sorted according to their ternary codes translated into decimal numbers. A remarkable gap is observed between the lowest linear stiffness of 1.16 N/mm where all inserts are removed and the highest limit of 4.37 N/mm where all holes are filled with longitudinal inserts. Meanwhile, selected patterns with only transverse inserts are tested as shown in Figure~\ref{fig3}a, featuring overlapped macroscopic behaviors in compression but programmability in the tension region instead. Furthermore, bidirectional stiffness programming with selected patterns involving mixed insert orientations displays bilinear constitutive relations as a superposition of the independent contributions by longitudinal and transverse inserts in the compression and tension regions, respectively, as illustrated in Figure~\ref{fig3}b.

\begin{figure}
  \includegraphics[width=\linewidth]{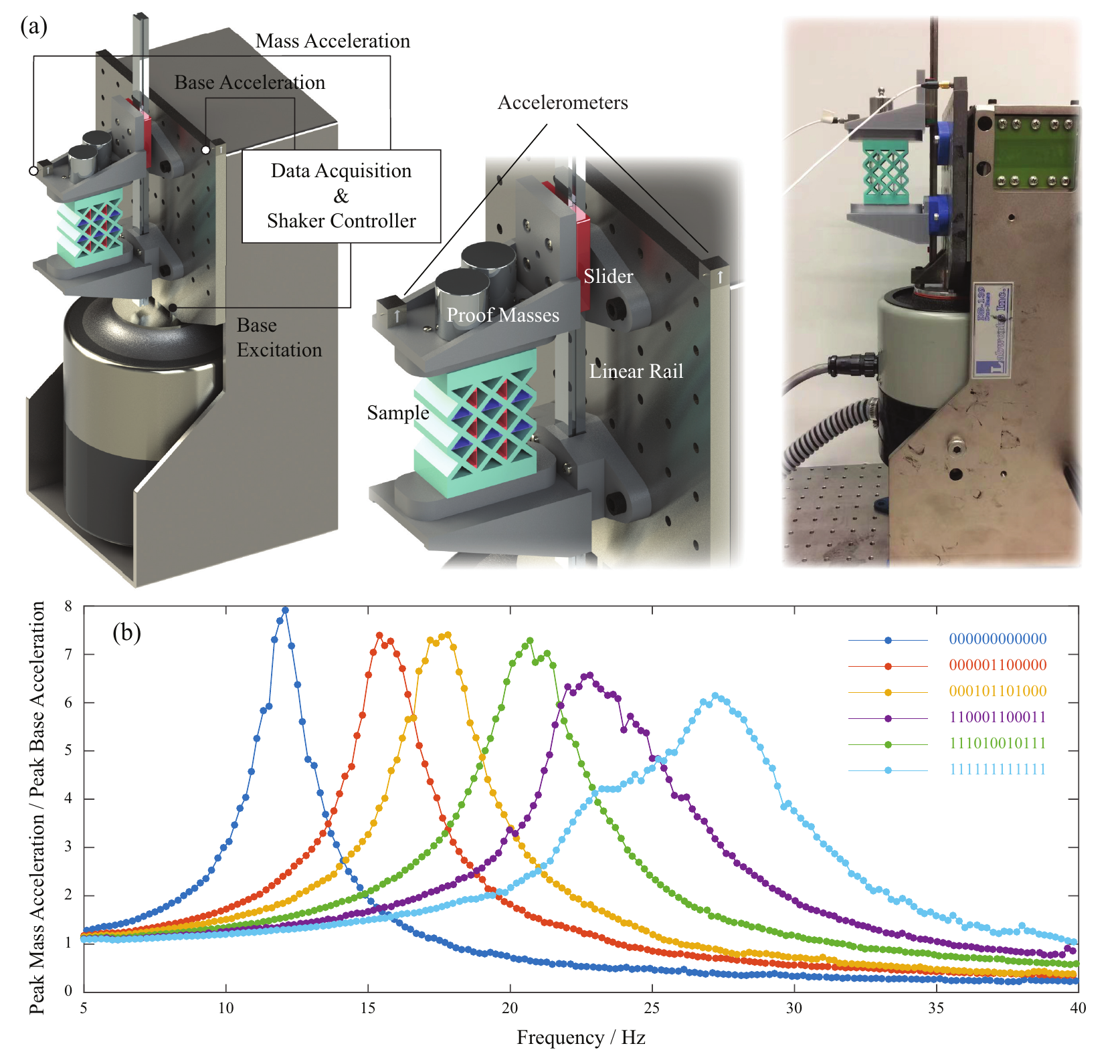}
  \caption{Prototype applied for vibration absorption.  a) Test setup. b) Experimental transfer functions from base to proof mass acceleration generated with sinusoidal dwell excitation series.}
  \label{fig4}
\end{figure}

Vast availability of programmable stiffnesses conveys applications in adaptive vibration isolation systems. When supporting an adequate mass upon a massive base, the prototype is anticipated to be capable of arbitrarily adjusting the system's resonant frequency up to doubling, given roughly the ratio between the upper and lower stiffness bounds observed in Figure~\ref{fig3}e and that the mass of inserts has a minimal effect. This is investigated on an electrodynamic shaker for selected patterns highlighted in Figure~\ref{fig2}b with only longitudinal inserts, as portrayed in Figure~\ref{fig4}a. The steady-state acceleration of a proof mass (0.2 kg), vertically connected via the programmed prototype onto the shaker base undergoing a series of sinusoidal excitations with constant magnitude 0.1$g$ (gravitational acceleration) dwelling at discrete frequencies, is recorded as in Figure~\ref{fig4}b. The static preload by gravity of the proof mass along with the low excitation magnitude ensures operation in the close-to-linear compression region. Results imply programmable resonant frequencies covering a broad band roughly from 12 Hz to 27 Hz, revealing potentials in rapid tuning of vibration isolation units in situations where resonance of base structures may shift in time.

\begin{figure}
	\includegraphics[width=\linewidth]{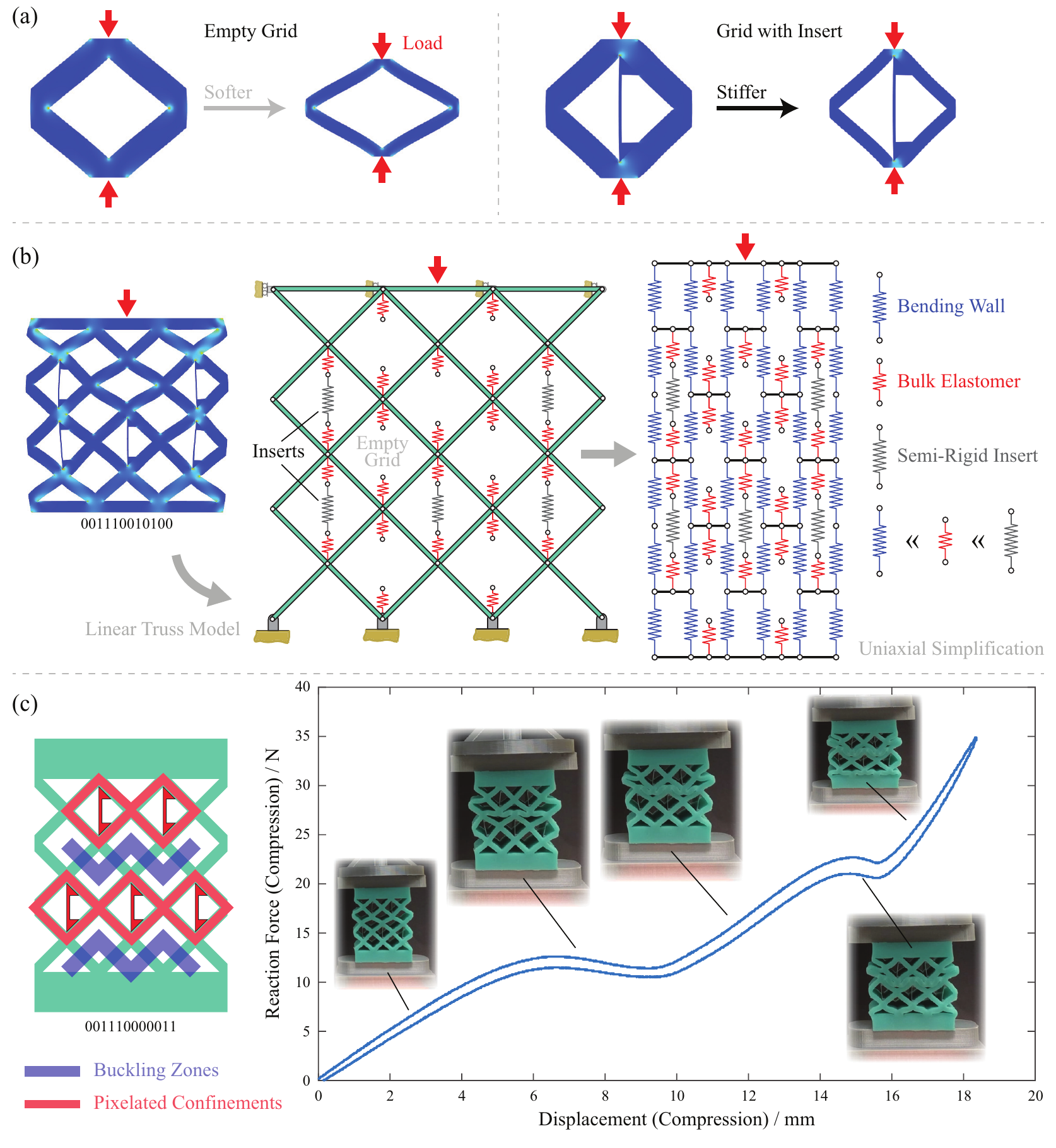}
	\caption{Interpretations of the fundamental mechanism. a) Improved programmable range by reducing elastomer thickness and b) the corresponding interpretation as series and parallel arrangements of switchable springs with an example pattern. c) Nonlinear piecewise stiffness programming with consecutive buckling zones.}
	\label{fig5}
\end{figure}

The above clarifies the proposed implementation of a stiffness programming strategy that achieves universal stability of geometrically defined input patterns and thus a direct encoding scheme, both static and in situ independent perturbation of unit cells in bidirectional operations, as well as a vast range of programmable stiffnesses. It is worth noting that such a ratio between upper and lower stiffness bounds can be extended without replacement of constituent materials but plainly by narrowing the elastomer matrix walls, as conceptually shown in Figure~\ref{fig5}a, which reduces the structural stiffness of a collapsible unit cell with no insert but meanwhile hardens one with an insert by cutting the volume of bulk elastomer around contact points. This accompanies a fundamental interpretation of the heterogeneous metamaterial near the linear region as a cascade of series and parallel arrangements of binary switchable springs which is a unidirectional simplification of a 2-dimensional truss model, both illustrated in Figure~\ref{fig5}b. Herein stiffnesses of three elements are distinctly ranked in the order of semi-rigid inserts, in-contact bulk elastomer and lastly the collapsible diamond elastomer grid, so that a series connection between the insert and the bulk elastomer triggers an equivalent stiffness of the latter, which further dominates a parallel combination with the equivalent stiffness of the collapsible grid. This model illuminates adaptations of the prototype as a platform for programming more sophisticated mechanical and dynamic behaviors by allowing freedom in the design of inserts for varieties of functions beyond rigid confinements. Meanwhile, explorations into nonlinear deformation regions may also be sought following a second interpretation of the design where zones of buckling beam arrays are indirectly programmed by constructions of opposite rigid confinements using inserts. Successive collapses and self contact of elastomer zones divided by confinements dominate the structure's behavior in different deformation phases, as demonstrated in Figure~\ref{fig5}c, which sees potentials in microscopic modifications of nonlinear constitutive relations of an extensive grid by programming clusters of unit cells as groups of pixelated confinement geometries.

\section{Conclusion}
Digital stiffness programming is realized using a 2-dimensional heterogeneous mechanical metamaterial featuring semi-rigid removable inserts to emulate both series and parallel arrangements of dissimilar springs, where each unit cell contributes uniquely to reach a resolution of linear stiffness programmability exponentially proportional to the number of unit cells. The design addresses common challenges in stiffness programming mechanisms and exhibits flexibilities as a platform for tuning complex macroscopic mechanical and dynamic behaviors by adopting various unit cell functions given the freedom of deploying any geometrically lockable functional insert.

\medskip
\noindent\textbf{Acknowledgements} \par
This work is supported by the National Science Foundation under Grant: CMMI 1662925. The authors would also like to acknowledge the financial support provided by the Purdue Research Foundation.

\bibliographystyle{MSP}
\bibliography{reference}
\newpage
\medskip

\clearpage

\medskip
\newpage
\pagestyle{empty}
\setcounter{section}{1}
\renewcommand\theequation{S.\arabic{equation}}
\section*{Supporting Information for Heterogeneous Digital Stiffness Programming}

\noindent \textit{Hongcheng Tao$^1$,~Francesco Danzi$^1$,~Christian E. Silva$^2$,~James M. Gibert$^1$}\\
\\
\noindent \textit{Ray W. Herrick Laboratories,\\
School of Mechanical Engineering,\\
Purdue University, West Lafayette, Indiana, 47907,USA}\\

\noindent \textit{Escuela Superior Polit\'ecnica del Litoral, ESPOL, \\
Km 30.5 Vía Perimetral, Guayaquil, Ecuador}\\

\renewcommand{\thefigure}{S\arabic{figure}}
\setcounter{figure}{0}

This section contains the Supplemental Information for  \textit{Heterogeneous Digital Stiffness Programming}.

\subsection{Programmable Patterns}
\begin{figure}
  \includegraphics[width=\linewidth]{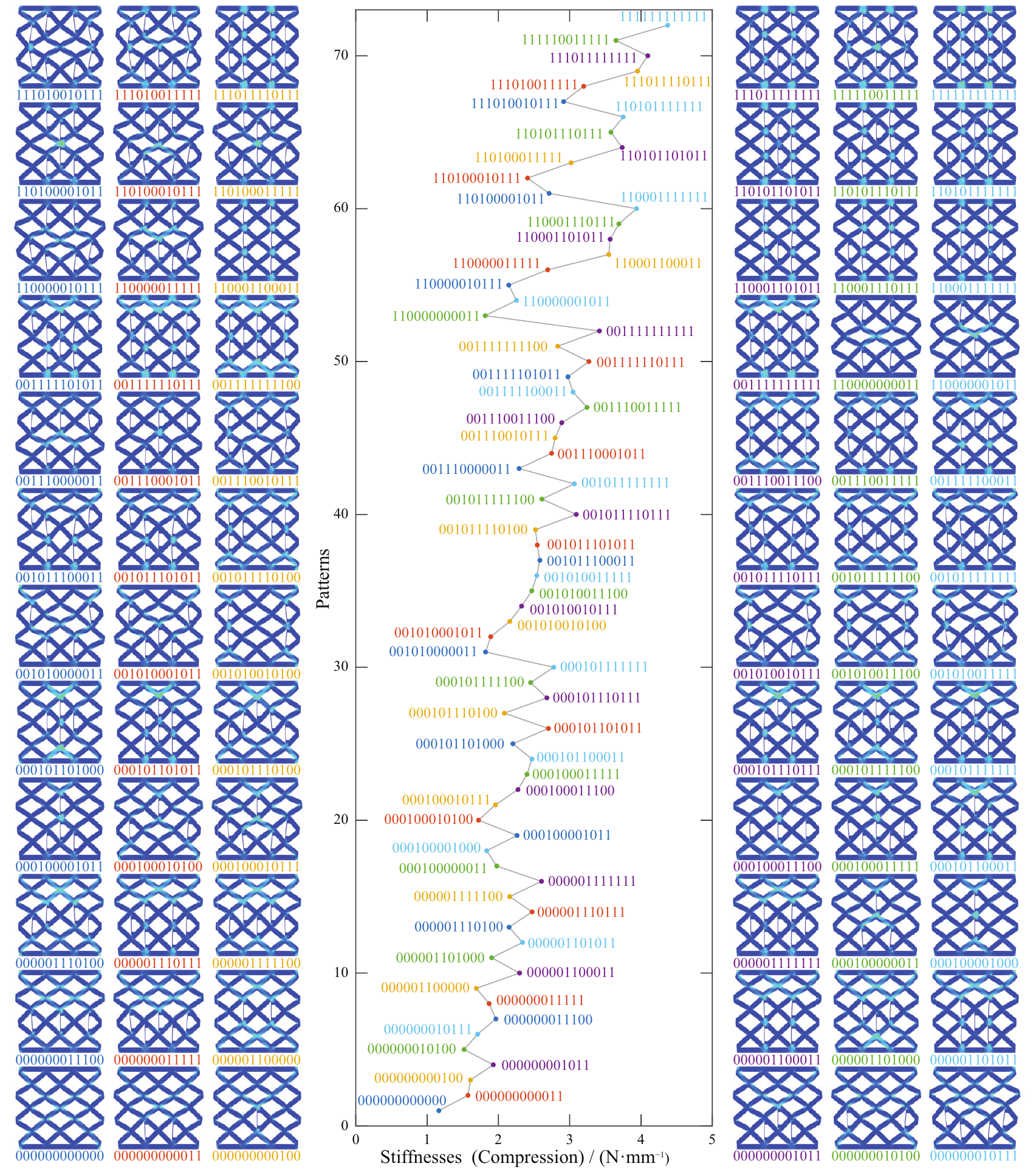}  
  \caption{Enumeration of all programmable patterns, ternary representation and their corresponding stiffness values. Note that the ternary representation in this case results in an equivalent binary number since the material is programmed for only compression.}
  \label{FigS0}
\end{figure}

 The 72 possible symmetric programmable patterns in compression, the ternary encoding, and their  corresponding stiffness values are shown in \textbf{Figure~\ref{FigS0}}.  Visualization of the patterns are shown through COMSOL finite element simulations that have contours plots  of the global strain superimposed on the structure. Note that if the material is programmed  in compression using longitudinal inserts, the states are restricted to ``0" and ``1" leading to a binary representation of the stiffness pattern. 
\newpage

\subsection{Uniaxial Approximation as Spring Network}

Stiffness programmability can be achieved by having an equivalent spring constant that has two discrete states depending on the presence of an inclusion that serves as a switch. When loaded in the linear range this results in $2^n$ unique stiffness values, where $n$ is the number of unit cells or digital elements. 

\begin{figure}[htb]
\centering
  \includegraphics[width=0.8\linewidth]{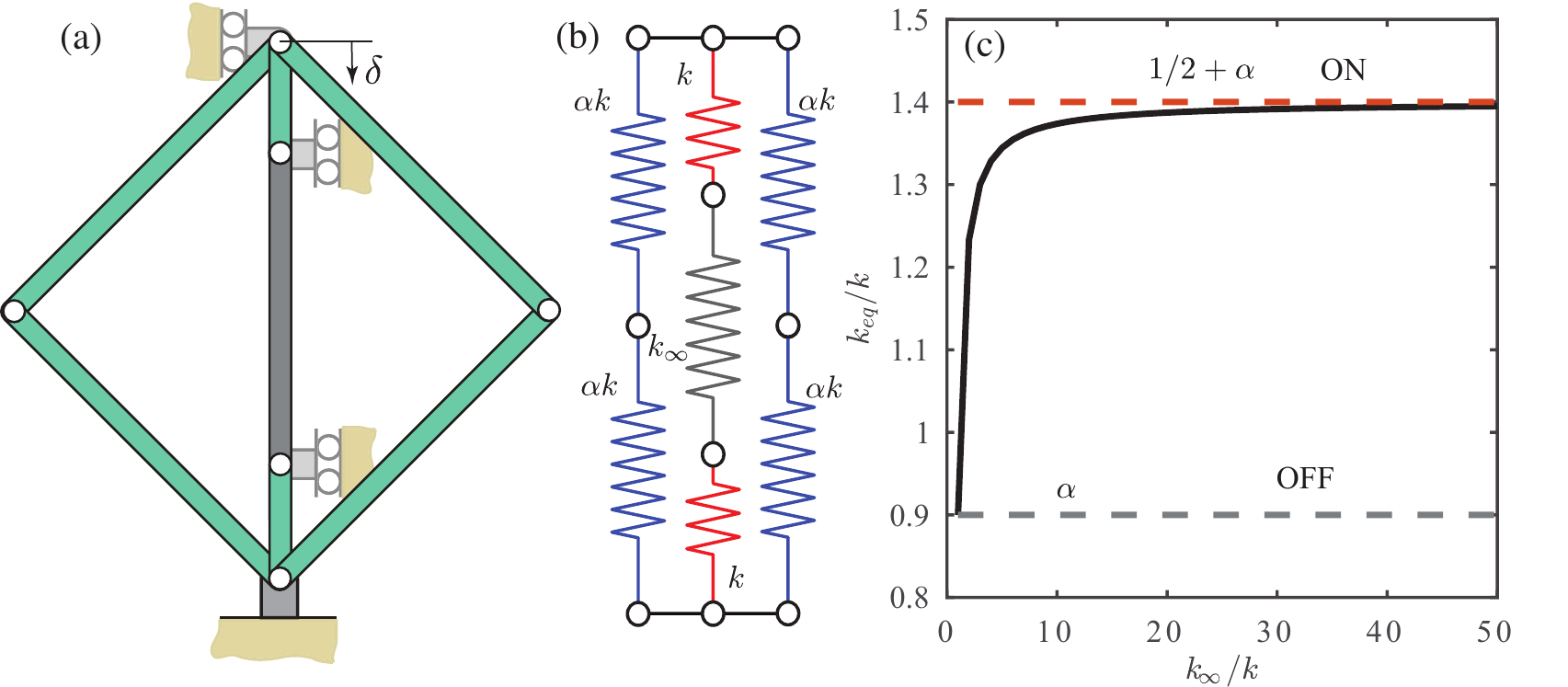}
  \caption{Behaviour of unit cell. a) Interpretation of the unit cell as a truss approximation. b) Simplification of the truss model as a uniaxial approximation.  c) Plot of equivalent stiffness ratio to insert stiffness ratio with bounds set as digital stiffness values. }
  \label{FigS1}
\end{figure}

In order to elucidate the mechanism of programmability, consider the digital stiffness element when the material is in compression.
A finite stiffness state  can be encoded in a material when it contains digital stiffness elements, $k^D_{eq}$, that behave as
\begin{align}
  k^{D}_{eq}(x)= \begin{cases}
\alpha k, &~\text{if} ~x =0~\text{(OFF)}, \\
(c+\alpha)k,&~\text{if} ~x =1~\text{(ON)},
\end{cases}\label{Eq1:BinaryStiff}  
\end{align}
where $c$ is constant and $\alpha$ is a constant between 0 and 1, and $k$ is a base stiffness value. Note that in tension the stiffness is  $k^{D}_{eq}(x)=\alpha k $.
The stiffness $k^{D}_{eq}(x)$  can be realized using  truss like structure shown in \textbf{Figure~\ref{FigS1}}a. The members of structure in green represents the elastomer and the member in grey represents a semi-rigid insert. This structure can be approximated by three columns of springs, \textbf{Figure~\ref{FigS1}}b.  The inner column has three springs in series with the  top and lower spring having a stiffness value of $k$ and middle spring has a  stiffness of $k_{\infty}$. The stiffnesses $k$ and $k_{\infty}$ represent the bulk elastomer, and semi-rigid insert, respectively. The outer columns represent the walls with two springs in series each having a stiffness, $\alpha k$. The equivalent stiffness of the system can be written as $k_{eq}=(\alpha k (2k_{\infty}+k)+kk_{\infty}) / (2k_{\infty}+k)$.  From this expression it becomes evident that the switch like behavior is dependent on $k_{\infty}$, representing the insert. 
Now, if $k_{\infty}=0$ then $k_{eq} \approx \alpha k$ and if $k_{\infty}\gg k$ then $k_{eq} \approx (1/2+\alpha)k$. 
 Plotting the ratio $k_{eq}/k$ versus $k_{\infty}/k$ shows that the stiffness ratio value starts at a value of $\alpha$ and asymptotically approaches $1/2+\alpha$,  \textbf{Figure~\ref{FigS1}}c, where  the ratio of $k_{\infty}/k\gg16$ provides a rudimentary bound to approximate the ideal digital stiffness element. 

\begin{figure}
\centering
  \includegraphics[width=.90\linewidth]{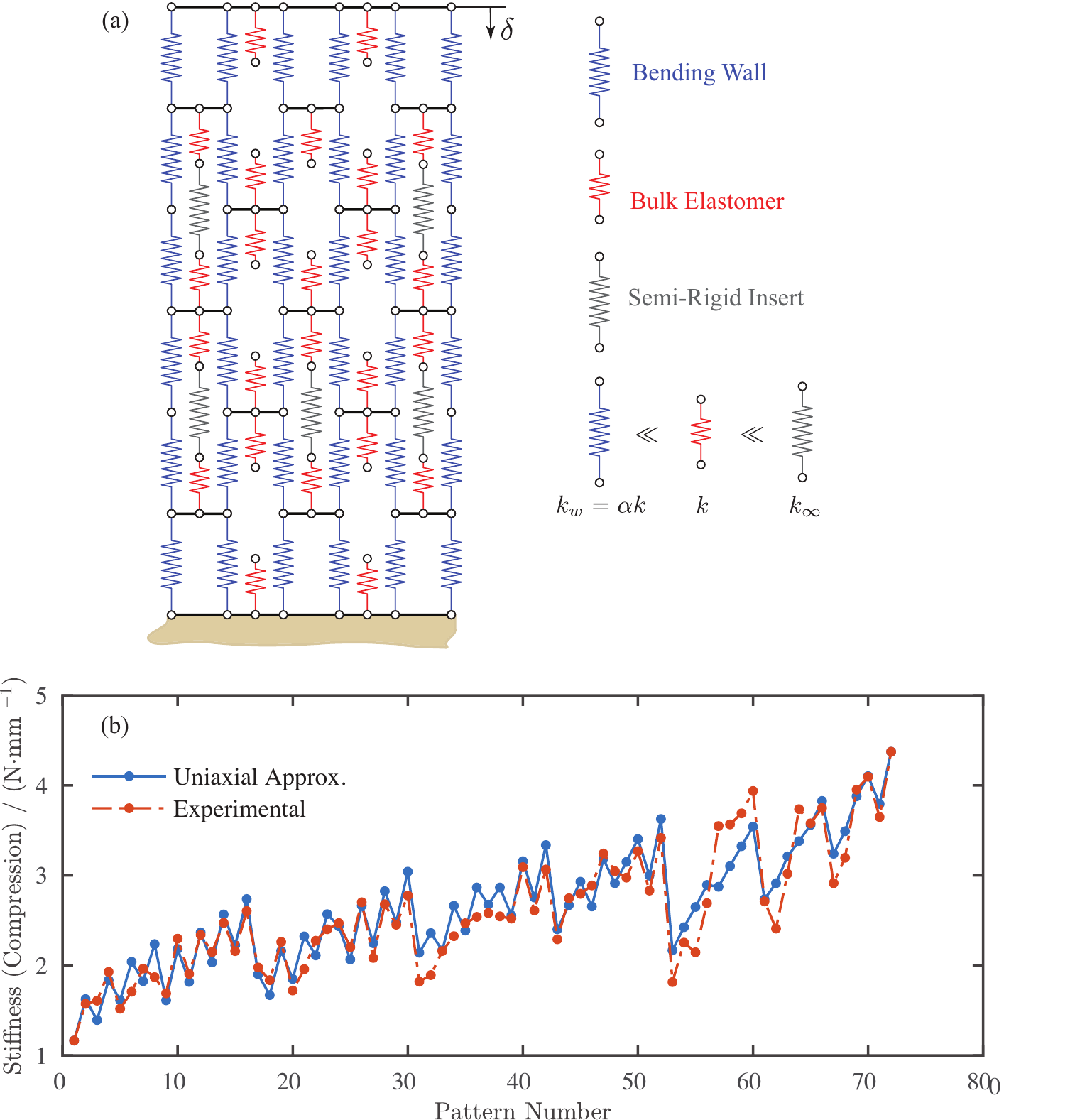}  
  \caption{Uniaxial simplification of heterogeneous metamaterial. a) Spring network model. b) Stiffness prediction from experimental analysis and uniaxial simplification.}
  \label{FigS:Uniaxial}
\end{figure}

 The programmable mechanical behavior can be illustrated using a network of springs representing  interconnected unit cells and connected walls, \textbf{Figure~\ref{FigS:Uniaxial}}a.
A material test can be simulated by enforcing the top link's deformation to have a prescribed displacement $\delta$ and calculating the reaction force, $F_R$. The system effectively acts as a spring with $F_R = k_{eff}\delta$ where $k_{eff}$ is the stiffness of the material if acted as a single spring. Comparing the predictions of stiffness from each of the 72 encodings of the spring network with the experimental stiffness. It is evident that the network closely captures the programmability of the prototype, \textbf{Figure~\ref{FigS:Uniaxial}}b.

\newpage

\subsection{Vibration Model}

\begin{figure}
  \centering
  \includegraphics[width=0.9\linewidth]{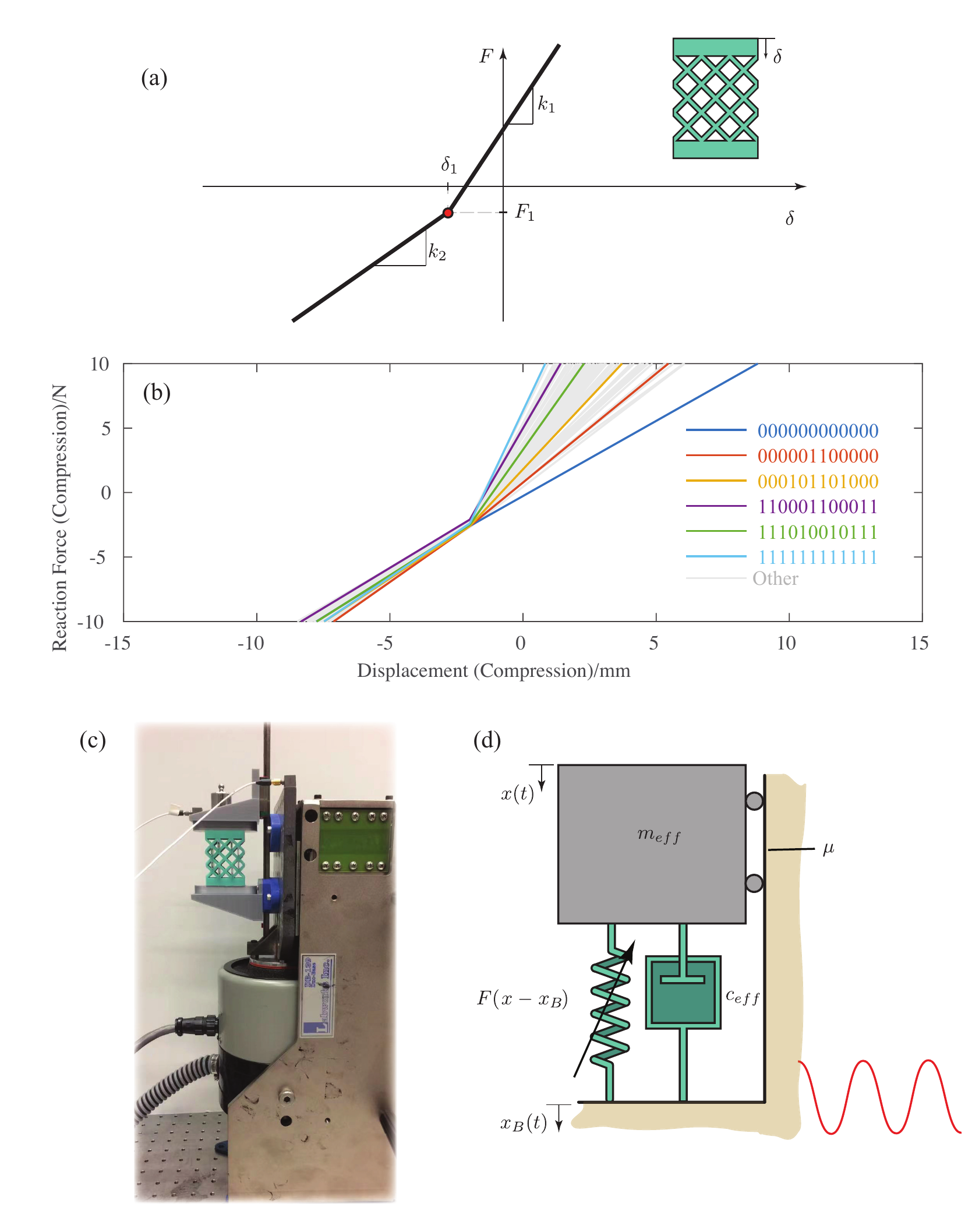}
  \caption{ Dynamic lumped parameter of metamaterial. a) Bilinear curve fit of all experimental data, selected curves are highlighted.  b) Pictorial description of bilinear force ($F$) displacement ($\delta$) plot. c) Test setup. d) Bilinear lumped parameter model.  }
  \label{FigS3}
\end{figure}

 The archetypal relationship for the heterogeneous metamaterial under small displacement with a digital element can be represented as the force displacement curve in  \textbf{Figure~\ref{FigS3}}a  and can be written as follows
\begin{equation}
F(\delta)= \begin{cases}
k_1\delta +F_1-k_1\delta_1, ~\text{if} ~\delta>\delta_1, \\
k_2\delta +F_1-k_2\delta_1,~\text{if} ~\delta<\delta_1,
\end{cases}\label{Eq1:BilinearStiff}
\end{equation}
where  $\delta$ is the stretch in the spring, $k_1$ is the  linear stiffness in compression, $k_2$ is the linear stiffness in tension, $F_1$  and $\delta_1$  are the corresponding force and displacement at the bifurcation point of the force displacement curve where the sitfness changes due to the presence of the insert.  The parameters are determined by experimentally curve fitting the compression data obtained by the Mark 10 compression testing frame. The linear fits from these tests are plotted in \textbf{Figure~\ref{FigS3}}b.

In order to understand the efficacy of the heterogeneous metamaterial for use in frequency programmability consider the vibration test shown in  \textbf{Figure~\ref{FigS3}}c and the corresponding model in \textbf{Figure~\ref{FigS3}}d.  The equation of motion of this system can be written as

\begin{subequations}
\begin{align}
&m_{eff}\ddot{\bar{\delta}}(t)+c_{eff}\dot{\bar{\delta}}(t)+\mu m_{eff}g\text{sgn}(\dot{\bar{\delta}}(t))+k_1\bar{\delta}(t)=k_1\delta_1+m_{eff}g-F_1-m_{eff}\ddot{x}_B(t),&~~\text{if}~\bar{\delta}(t)>\delta_1,
\label{S3a}\\
&m_{eff}\ddot{\bar{\delta}}(t)+c_{eff}\dot{\bar{\delta}}(t)+\mu m_{eff}g\text{sgn}(\dot{\bar{\delta}}(t))+k_2\bar{\delta}(t)=k_2\delta_1+m_{eff}g-F_1-m_{eff}\ddot{x}_B(t),&~~\text{if}~\bar{\delta}(t)<\delta_1,
\label{S3b}
\end{align}
\end{subequations}
where $\bar{\delta}(t)=x(t)-x_B(t)$, is defined as the compression in the material, ~$x(t)$ is the absolute motion of the proof mass,~$\ddot{x}_B(t)$ is the base acceleration defined as   $\ddot{x}_B(t)= G\sin\omega t$, $G$ is the amplitude of acceleration, $\omega$ is the frequency of excitation, $m_{eff}$ is the sum of the mass of the insert, equivalent mass  of  the elastomer, and the proof mass, $c_{eff}$ is damping coefficient of the material, $g$ is the gravitational constant. The coefficient between the components of the rail guide is denoted as $\mu$. The effective mass and stiffness  in tension are adjusted from their nominal values.  This was done to: 1)  match the experimental resonant frequency, and 2) account for the inserts slightly losing contact when the elastomer vibrates when the structure is in tension. 
The changes in these parameters along with  coefficient of friction, and damping coefficient are determined from fitting two experimental frequency response function  (FRF) curves.  The Equations
~\ref{S3a} and~\ref{S3b} are solved numerically in Matlab using the solver ODE45. The FRF curves were obtained by vary the forcing frequency and solving the equations of motion for each frequency until the system reached steady state. At steady state the peak amplitudes of response were recorded.

\begin{figure}
  \includegraphics[width=0.9\linewidth]{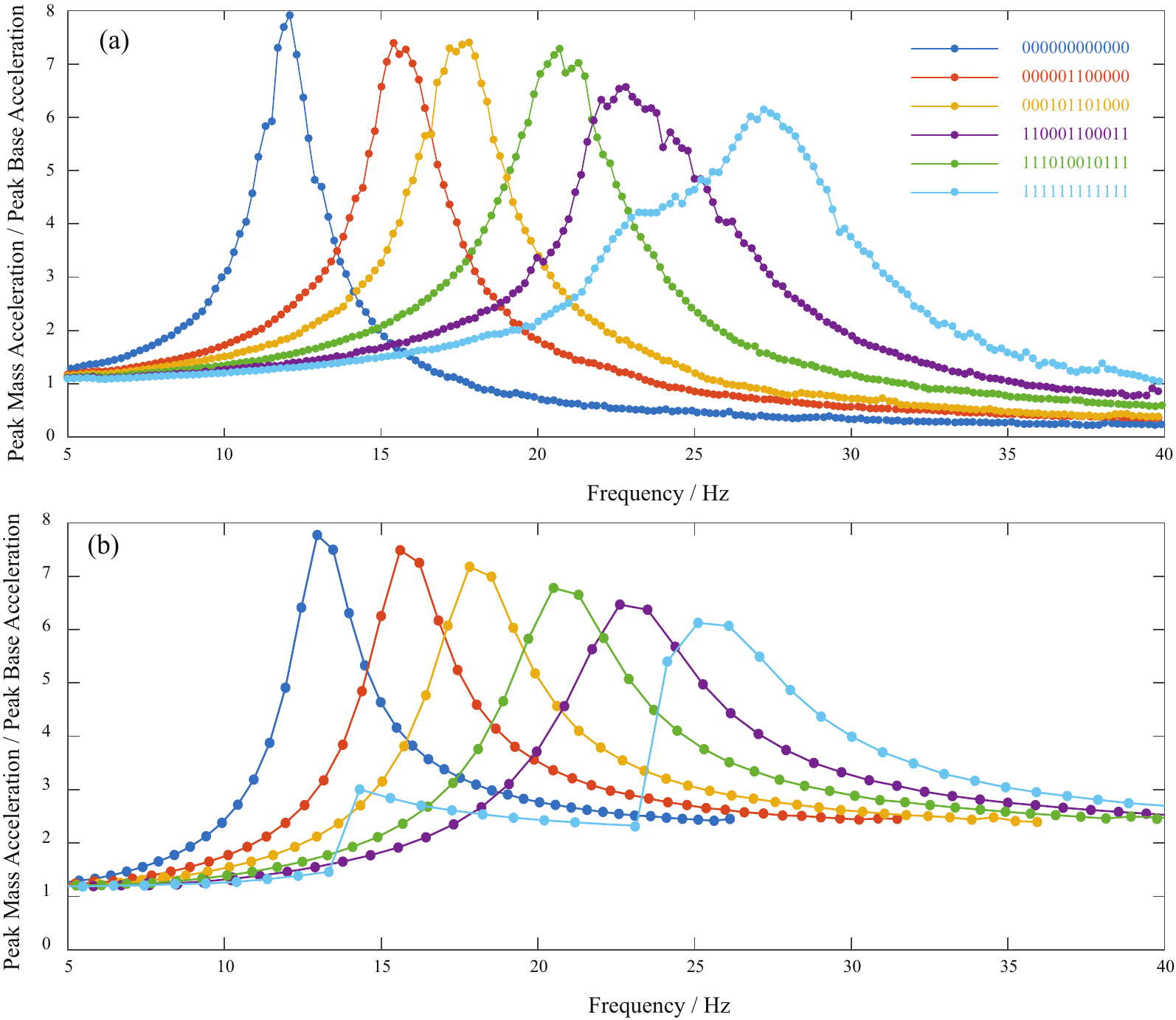}  
  \caption{Transfer functions. a) Experimental transfer functions from base to proof mass acceleration. b) Theoretical transfer functions from base to proof mass acceleration.}
  \label{FigS4}
\end{figure}

\textbf{Figure~\ref{FigS4}}a is a theoretical plot of the frequency response function between the base acceleration to the proof mass acceleration.
For all configurations tests the theoretical responses agree qualitatively to the experimental ones in  \textbf{Figure~\ref{FigS4}}b. The  magnitudes of vibration and the resonant peaks have similar values. Note that in both the experimental case and in the model given in  Equations~\ref{S3a} and~\ref{S3b} the FRF for configuration 111111111111 has a distorted resonant peak.  The model captured the presence of the distortion but  does not match the shape of the FRF curve. In this configuration, the model indicated that the distortion occurred over a limited frequency range and was due to the metamaterial switching from vibrating strictly below $\delta_1$, to vibrating in a region above and below
$\delta_1$.

\clearpage

\end{document}